\documentclass[journal]{IEEEtran}
\usepackage{amsmath,amsfonts,color}
\usepackage{algorithmic}
\usepackage{algorithm}
\usepackage{array}
\usepackage{lipsum}
\usepackage[caption=false,font=normalsize,labelfont=sf,textfont=sf]{subfig}
\usepackage{textcomp}
\usepackage{stfloats}
\usepackage{url}
\usepackage{verbatim}
\usepackage{graphicx}
\usepackage{cite}
\def\BibTeX{{\rm B\kern-.05em{\sc i\kern-.025em b}\kern-.08em
    T\kern-.1667em\lower.7ex\hbox{E}\kern-.125emX}}

\newcommand{\E}{\mathbb{E}} 
\newcommand{\correction}[1]{{#1}}

\begin{document}

\title{Second-Order Characterization of Micro Doppler Radar Signatures of Drone Swarms}
\author{Anders Malthe Westerkam, Alba Spliid Damkjær, Rasmus Erik Villadsen, Magnus Ørum Bastrup Poulsen, \\Troels Pedersen. \\Aalborg University, Aalborg Denmark. Email: \{amw,troels\}@es.aau.dk
\thanks{This work is funded by the Thomas B. Thriges Foundation grant 7538-1806.
}

}
\maketitle

\begin{abstract}
We propose an analytic model for the second-order characteristics of the radar return signal from a swarm of rotor drones. We consider the case of a swarm of identical drones, with each a number of rotors comprised of a number of rotor blades. By considering the orientation and speed of each rotor as stochastic variables, we derive expressions for the autocorrelation function (ACF) and power spectral density (PSD). The ACF and PSD are in the form of an infinite series with coefficients that drop to zero at a predictable limit. Thus in practical applications, the series may be truncated. As a special case, we  show that for deterministic rotor speed, the ACF can be expressed in closed form. We further investigate how system parameters (Blade length, Rotor speed, number of blades, and number of drones) influence the derived expressions for the ACF and PSD.
\end{abstract}


\section{Introduction}

The advancement and widespread use of micro unmanned aerial vehicles (UAVs), commonly known as drones, have increased the need for effective detection and classification techniques, particularly within restricted airspace \cite{Pothana2023}. In addition to single drone usage, multiple drones flying simultaneously in close proximity, referred to as drone swarms, have also seen gradual deployment for various malicious purposes \cite{Liu2023}. Typical rotor drones, with their small radar cross section, slow speeds, and possible low flight altitudes, present new challenges for radar detection compared to more traditional radar targets.  The presence of birds further complicates the methods for detection due to their often similar size. 

Recent advances show that it is possible to distinguish rotor drones and birds as they produce different micro-Doppler signatures \cite{Liu2021}. Some works focus on measurement-driven micro-Doppler characterisation for single drones \cite{Lehmann2020,Leonardi2022,Bujakovic2022}. Others have considered computer-intensive deterministic models for the radar return signal \cite{Chen2006,Cai2019,Costa2024}. Neither of these approaches is straightforward to use for analysis. In particular, new measurements or simulations should be performed to answer questions regarding how different drones and swarm configurations affect properties of the micro-Doppler signature. Some work has been done in expressing the micro-Doppler signal arising from propellers \cite{Martin1990}, however, from these models, it is still difficult to infer how movement of drone swarms affects the micro-Doppler response.


{In the present contribution, we construct an analytic stochastic model for the micro-Doppler signature of a swarm of small rotor drones observed by a monostatic radar. We aim to obtain a model that directly shows the effect of key drone parameters on the autocorrelation function (ACF) and power spectral density (PSD) of the micro-Doppler signature. Therefore, we purposely make crude assumptions that ignore the effects present in the real scenario but would obscure the analysis. The list of ignored effects includes: macro-Doppler, clutter, noise, tilt of the drone, blade-flashes, multipath, \correction{hardware limitations}, and shadowing by the drone body. Furthermore, all drones are assumed to be of the same type. Nevertheless, the proposed model is adjustable in the number of drones, number of rotors per drone, number of blades per rotor, blade length, and rotor speed and accounts for the wavelength of the radar signal. Obviously, greater realism may be obtained by accounting for more effects, at the cost of mathematical complexity and possibly reduced interpretability. This approach allows for the derivation of both the ACF and PSD in closed form. Numerical examples are included to illustrate the derived expressions. \correction{The model herein could possibly be leveraged in, e.g. automatic drone swarm detection, interference mitigation, or target classification for real-time radar systems.}} 

\begin{figure}
    \centering
    \includegraphics[page = 1,width=0.475\textwidth]{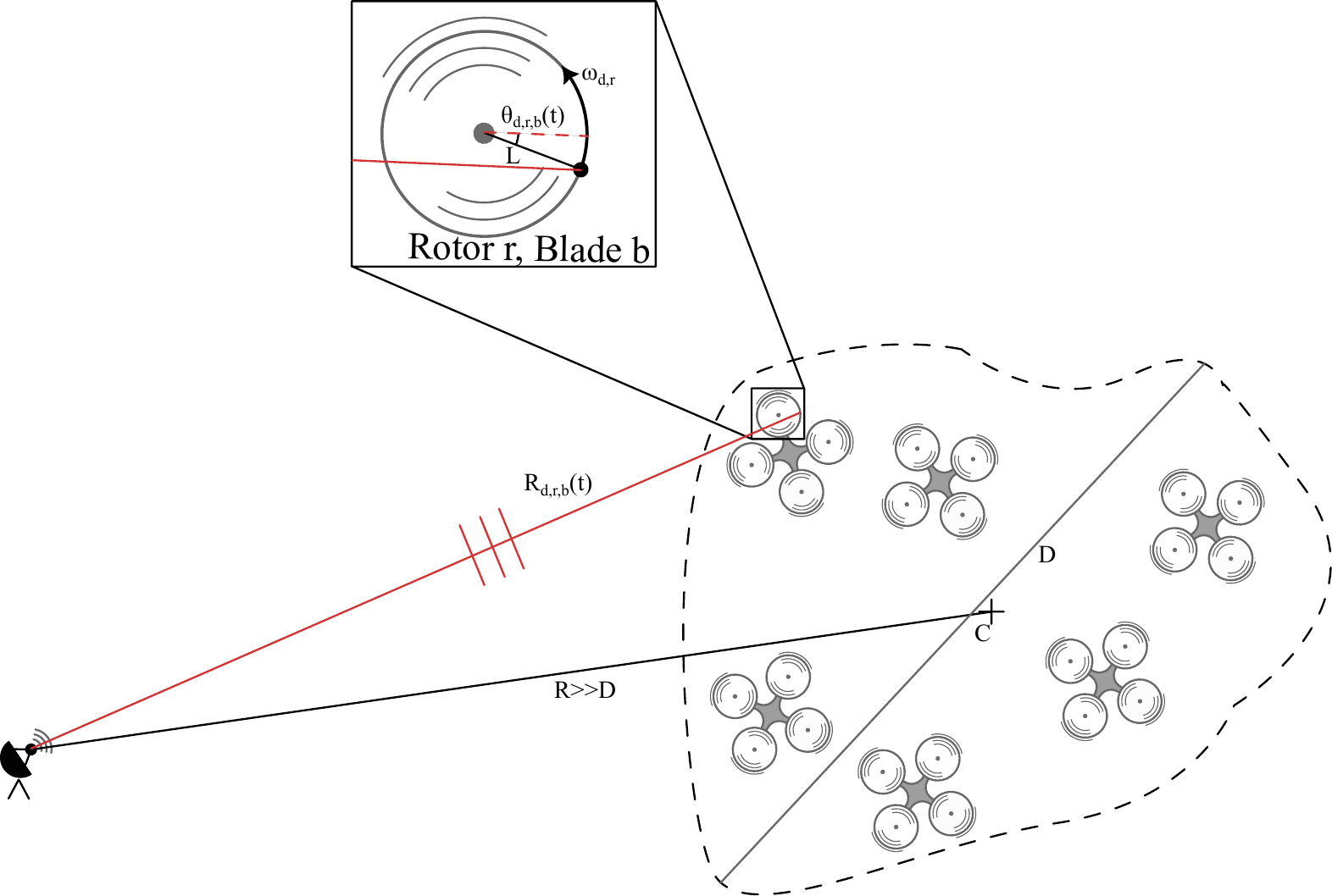}
    \caption{Monostatic radar at range R from the centre C of a drone swarm situated in a region of diameter D much smaller than R. Each drone in the swarm has $N_r$ rotors with $N_b$ blades.}\label{fig:drone_swarm}
\end{figure}

\IEEEpubidadjcol

\section{Signal Model}

Consider the scenario illustrated in Fig.~\ref{fig:drone_swarm} with a drone swarm of diameter $D$ at a range $R$ observed by a monostatic pulsating radar. The swarm consists of $N_d$ identical drones, each with $N_r$ rotors and $N_b$ blades of length $L$ on each rotor. Now, assume that $R\gg D$, such that the drone swarm is in the far field of the radar. Let the carrier wavelength $\lambda\ll L$ and assume that the signal bandwidth is small enough so that drones become indistinguishable in delay. Furthermore, $R$ is assumed much larger than the flying altitude of the swarm, allowing a 2D approximation of the scene. 

We focus on the micro-Doppler signature of the drones and therefore omit macro-Doppler, noise, and clutter in the model. If needed, these effects may be included to obtain a more comprehensive modelling effort or simulated studies. 
Assuming ideal back scatter conditions, we then reduce all drones to individual sets of spinning point scatterers, each scatterer with range $R_{d,r,b}$ to the radar. 

The angular velocity of rotor $r$ on drone $d$ is denoted by $\omega_{d,r}$, and the angle of blade $b$ by,
\begin{equation}
    \theta_{d,r,b}(t)=\theta_{d,r}+\omega_{d,r}t+2\pi \frac{b-1}{N_b},
\end{equation}
where $\theta_{d,r}$ is the initial angle distributed as $\{\theta_{d,r}\}_{d,r\in\mathbb{N}_0}  \stackrel{i.i.d.}{\sim}U(0,2\pi)$. Lastly, let $g$ be the amplitude and reflection phase shift of the returned signal, here assumed identical for all blades, then the signal model is 

\begin{equation}\label{eq:return_approx}
    y(t)= g\sum_{d}^{N_d}\sum_r^{N_r}\underbrace{{e}^{-i\gamma_{d,r}}}_{h_{d,r}}\sum_b^{N_b}{e}^{-i\frac{4\pi}{\lambda}L\cos(\theta_{d,r,b}(t))},
\end{equation}
where $\gamma_{d,r}$ is the change in phase when the line $R_{d,r,b}$ is projected onto the line $R$. The returned signal model becomes stochastic by assuming $\{\gamma_{d,r}\}_{d,r\in\mathbb{N}_0} \stackrel{i.i.d.}{\sim}U(0,2\pi)$, and $\{\omega_{d,r}\}_{d,r\in\mathbb{N}_0}  \stackrel{i.i.d.}{\sim}\mathcal{N}(\Bar{\omega},\sigma^2_{\omega})$, with $\bar{\omega}$ being the mean and $\sigma_{\omega}^2$ being the variance. Since $y(t)$ is symmetric in $\omega_{d,r}$, even though rotors spin in both directions, it suffices to consider only positive rotation. {We model the angular velocity independently between all rotors in the swarm, accounting for different rotor speeds which may arise due to, e.g., differing air density or manoeuvring. Note that modelling all angular velocities with independent Gaussians with different means and variances is a straightforward extension.} 

\section{First- and Second-order properties}\label{sec:Second_order_properties}
The mean of the signal $y(t)$ in \eqref{eq:return_approx} is zero as $\E[h_{d,r}] = 0$. Thus, the covariance function equals the ACF  
\begin{equation}\label{eq:ACF_first}
    \mathcal{R}_s(t,\tau)=\E\left[y(t)y^*(t+\tau)\right].
\end{equation}
Since $\E[h_{d,r}h^*_{d',r'}] = \delta_{d,d'}\delta_{r,r'}$, where $\delta$ is the Kronecker delta, we obtain
\begin{equation}            
    \mathcal{R}_s(t,\tau) = |g|^2\sum_{d,r}R_{rotor}(t,\tau)
\end{equation}
with the ACF for a single rotor defined as,
\begin{equation}\label{eq:expectation_over_outer_product}
    \mathcal{R}_{rotor}(t,\tau)= \sum_{b,b'}\E\left[{e}^{il\sin\left(\frac{\tau\omega_{d,r}}{2}-\pi\frac{b-b'}{N_b}\right)\sin\left(\phi(t,\tau)\right)}\right].
\end{equation}
Here, $l = 8\pi L/\lambda$, and $\phi(t,\tau) =\theta_{d,r,b}(t)+\theta_{d,r,b'}(t+\tau)=\theta_{d,r}+\frac{\omega_{d,r}(2t+\tau)}{2}+\pi\frac{b-b'-2}{N_b}$. 
Considering the phase $\phi(t,\tau)$ modulo $2\pi$, it is  straightforward to show that it is uniformly distributed on $[0,2\pi)$ for all $t$ and $\tau$. Thus, signal of a single rotor is a wide sense stationary process.  Expanding \eqref{eq:expectation_over_outer_product} using the Jacobi-Anger expansion, and taking the expectation with respect to $\omega_{d,r}$ yields
\begin{multline}\label{eq:first_expectation_taken}
\mathcal{R}_{rotor}(\tau)=\\\sum_{b,b'}\sum_{n=-\infty}^{\infty} \E_\phi[J_n(l\text{sin}(\phi))]e^{i\left(\frac{\Bar{\omega}\tau}{2}-\pi\frac{b-b'}{N_b}\right) n}e^{-\frac{\sigma^2_{\omega}\tau^2n^2}{8}},
\end{multline}
where $J_n(\cdot)$ denotes the Bessel function of the first kind, order $n$. By the result in Appendix \ref{app:Derivation_of_series_coefficients},

\begin{multline}
    \mathcal{R}(\tau) = |g|^2N_dN_rN_b^2 J_0^2\left(\frac{l}{2}\right) \\+ N_rN_d\sum_{n=1}^{\infty}K_n\text{cos}(n\Bar{\omega}\tau)e^{-\frac{\sigma^2_{\omega}\tau^2n^2}{2}},
\end{multline}
with
\begin{equation}
    K_n = N_b\left[1+2\sum_{m=1}^{N_b-1}\left(1-\frac{m}{N_b}\right)\text{cos}\left(2\pi n \frac{m}{N_b}\right)\right ]J_n^2\left(\frac{l}{2}\right).
\end{equation}
Noting that the bracket equals $N_b$ when $n$ is a multiple of $N_b$ and zero otherwise, we have
\begin{multline}\label{eq:ACF_final}
    \mathcal{R}_s(\tau) = |g|^2N_dN_rN_b^2\times  \\ \left[J_0^2\left(\frac{l}{2}\right) + 2\sum_{n=1}^{\infty}J_{N_b n}^2\left(\frac{l}{2}\right)\text{cos}(N_b n\Bar{\omega}\tau)e^{-\frac{\sigma^2_{\omega}\tau^2n^2N_b^2}{2}}\right].
\end{multline}

To analyse the effect of the physical parameters in the ACF, starting with the coefficients $J_{N_bn}^2(l/2)$. We note that $J_\nu^2(z) \approx 0$ before its first peak. From Abramowitz and Stegun (9.5.16)\cite{abramowitz+stegun} we see that the first zero of the Bessel function and its derivative to the first order is $z \approx \nu$, hence we have
\begin{equation}
    \frac{d}{d z} (J_\nu(z))^2 = 0 \rightarrow z\approx \nu,
\end{equation}
from which we truncate the sum in \eqref{eq:ACF_final} by,
\begin{equation}\label{eq:upper_limit}
    J_{nN_b}^2\left(\frac{l}{2}\right) \approx 0,\phantom{m} n > \frac{l}{2N_b}.
\end{equation}

To consider the special case with $\omega_{d,r}$ deterministic ($\sigma_{\omega_{r,d}}=0$), we start from \eqref{eq:expectation_over_outer_product}, and expand using Jacobi-Anger as,
\begin{equation}
    \mathcal{R}_{rotor}(\tau) = \sum_{b,b'}\sum_{n=-\infty}^\infty J_n\left(l\sin\left(\frac{\Bar{\omega}\tau}{2}-\pi\frac{b-b'}{N_b}\right)\right)\E\left[e^{in\phi}\right].
\end{equation}
The expectation is nonzero only at $n=0$ and thus,
\begin{align}\label{eq:ACF_no_omega}
    \mathcal{R}_{\sigma_{\omega}=0}(\tau)=|g|^2N_dN_r\sum^{N_b}_{b, b'}J_0\left(l\sin\left(\frac{\bar{\omega}\tau}{2}-\pi\frac{b-b'}{N_b}\right)\right). 
\end{align}

The main lobe width is obtained by considering the limit $\tau\rightarrow0$. Noting that the sum in \eqref{eq:ACF_final} is truncated by the result in \eqref{eq:upper_limit}, we see that the exponential in the ACF is unity for all $n$ considered. This is equivalent to the case $\sigma_{\omega}=0$, and hence the result obtained in \eqref{eq:ACF_no_omega}. Employing the small angle approximation for the sine, and considering the zeros of the Bessel function, yields the following approximate width of the main lobe,
\begin{equation}
    \Delta\tau_{\text{null-to-null}} \approx \frac{4.8}{l\bar{\omega}}.
\end{equation}

\begin{figure}
    \centering
    \includegraphics[width=1\linewidth]{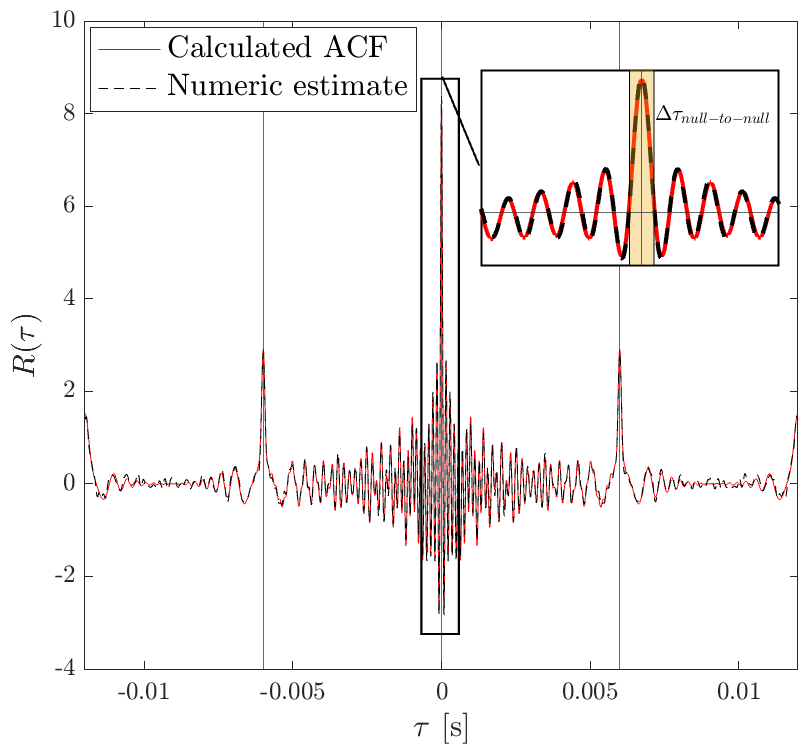}
    \caption{The ACF with parameter setting, in Fig. \ref{fig:spectogram}}   
    \label{fig:ACF_figure}
\end{figure}
This range is also shown in the insert of Fig.~\ref{fig:ACF_figure}. Note that the main lobe width is inversely proportional to $l\bar{\omega}$.

\begin{figure}
    \centering
    \includegraphics[width=1\linewidth]{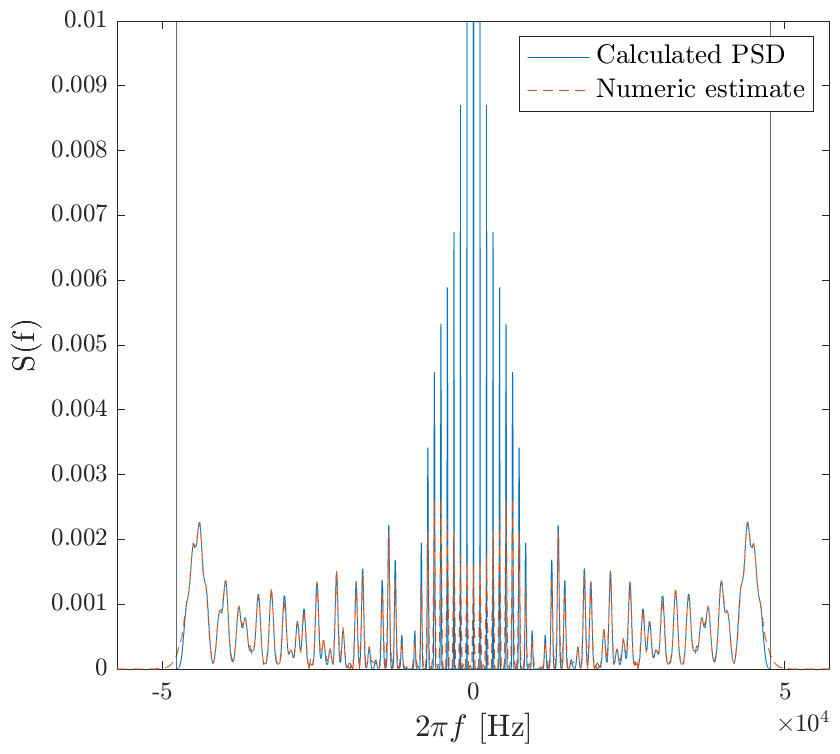}
    \caption{The PSD for a drone modelled after the DJI Mavic. The vertical lines denote the range in which the PSD is calculated for the parameters in Fig.~\ref{fig:spectogram}}
    \label{fig:PSD_figure}
\end{figure}

The PSD is derived from \eqref{eq:ACF_final} by Fourier transform, The summands in \eqref{eq:ACF_final} are a product of two functions, with Fourier transforms,
\begin{equation}
    \mathcal{F}\left\{\text{cos}(N_b n \Bar{\omega}\tau)\right\} = \sqrt{\frac{\pi}{2}}(\delta(f+N_bn\Bar{\omega})+\delta(f-N_bn\Bar{\omega})),
\end{equation}
\begin{equation}
    \mathcal{F}\left\{e^{-\frac{\sigma^2_{\omega}\tau^2n^2N_b^2}{2}}\right\} = \frac{e^{-\frac{f^2}{2\sigma^2_{\omega}n^2N_b^2}}}{\sqrt{\sigma^2_{\omega}n^2N_b^2}}.
\end{equation}
By the convolution theorem we obtain the PSD as,
\begin{multline}\label{eq:PSD_final}
    S(f) = \sqrt{2\pi}|g|^2N_dN_rN_b^2 J_0^2\left(\frac{l}{2}\right) \delta(f) + |g|^2\frac{\sqrt{2\pi}N_b}{\sigma_{\omega}}\\\times N_dN_r\sum_{n=1}^{\infty}\frac{J_{N_bn}^2\left(\frac{l}{2}\right)}{n}\left[e^{-\frac{(f-N_bn\bar{\omega})^2}{2\sigma^2_{\omega}n^2N_b^2}}+e^{-\frac{(f+N_bn\bar{\omega})^2}{2\sigma^2_{\omega}n^2N_b^2}}\right].
\end{multline}
This is a sum of Gaussian kernels placed symmetrically around zero. Due to Gaussian symmetry it is also possible to predetermine the range of frequencies which are of interest based on \eqref{eq:upper_limit} as the furthest Gaussian is placed at $\pm \Bar{\omega} l/2$, and as it has standard deviation $\sigma = \sigma_{\omega}l/2$, we define the interval as,
\begin{equation}
    \Delta f = \frac{l}{2}\left[-\bar{\omega}-5\sigma,\bar{\omega}+5\sigma\right].
\end{equation}
Here, we use five "standard deviations" as it effectively captures the whole tail. The PSD is shown calculated in this interval in Fig.~\ref{fig:PSD_figure}.

We remark that in the limit $\sigma_\omega \rightarrow 0$ the kernels in \eqref{eq:PSD_final} approach Dirac deltas, and we recover the conclusions from \cite{Martin1990}, with the frequency components separated by $N_b\Bar{\omega}$ symmetrically around zero, with a total bandwidth of $B=l\Bar{\omega}$, and the total number of components equal to $l/N_b$. 

\begin{figure}
    \centering
    \includegraphics[width=1\linewidth]{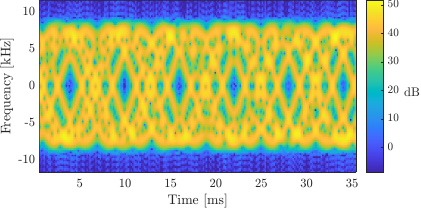}
    \caption{Spectrogram of the signal model with parameters, $L = 21$ cm, $\lambda = 3$ cm, $N_b = 2$, $N_r = 4$, $N_d = 1$, $\Bar{\omega} = 523$ $s^{-1}$, $\sigma_\omega^2 = 27$ $s^{-2}$}
    \label{fig:spectogram}
\end{figure}

\section{Numeric analysis}
We compare the results derived in \correction{Section}~\ref{sec:Second_order_properties} against a numeric estimate of the ACF and PSD meant to resemble a commercial drone.  The parameters are listed in Fig.~\ref{fig:spectogram}.
We generate 50\,000 realizations of signals according to \eqref{eq:return_approx} with 4001 time samples, of which the short-time Fourier transform is shown in Fig.~\ref{fig:spectogram}. By visual inspection, this realisation resembles spectrograms reported in \cite{Lehmann2020}, however, without the "blade flashes" characteristic of when the blade is orthogonal to the radar. 

The theoretical ACF and PSD are compared to those obtained from $N$ simulation runs as,
\begin{equation}
    \hat{R}(\tau) = \frac{1}{N}\sum_{n=1}^NS_n(t)S_n^*(t+\tau), \phantom{mm} \hat{S}(f) = \text{FFT}\{\hat{R}(\tau)\}.
\end{equation}
Fig.~\ref{fig:ACF_figure} shows good agreement between the calculated ACF and the numerical estimate over the whole time delay. Also,  the PSD in Fig.~\ref{fig:PSD_figure} agrees well with the numerical estimate. The minor discrepancies in peak height for the PSD at low frequencies are attributed to how narrow the Gaussian kernels are in this region according to \eqref{eq:PSD_final}.  
\begin{figure}
    \centering
    \includegraphics[width=1\linewidth]{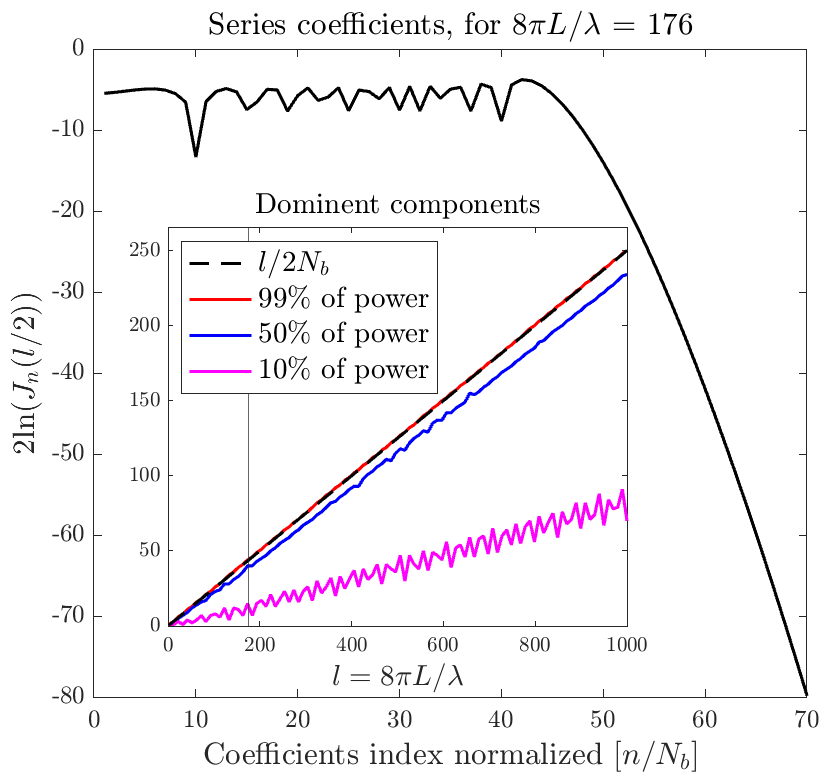}
    \caption{Value of the coefficients for the settings considered in this article. Further, the insert shows a comparison of how many coefficients are needed to contain a percentage of the power within the first 10\,000 coefficients for a range of $l$ values.}
    \label{fig:coeeficient_figure}
\end{figure}

It is worth considering whether all components up to the limit defined in \eqref{eq:upper_limit} are equally important. To investigate this, we plot the $\log$ of the coefficients for a drone swarm configuration as shown in Fig.~\ref{fig:coeeficient_figure}. Based on the limit, we would expect the power of the coefficients to sharply drop off after index 44, which is also observed. Furthermore, it can be seen that, prior to the limit, the values seem to be similar in strength. This is a general property, as is shown in the insert, which shows the number of coefficients as a function of $l$. Here, it can be seen that the power of the series is spread across the coefficient number. Hence, to derive a rule for selecting the most important coefficients is not straightforward.

\section{Conclusion}
The autocorrelation function and power spectral density of the micro-Doppler signal of a drone swarm were calculated under simplifying assumptions. Under these assumptions, interpretable results are obtained, e.g., linking the width of the main lobe to the system parameters of the drone, as well as the frequency range of the power spectral density. 
\correction{These results help to understand how micro-Doppler signatures reflect drone parameters and which information can potentially be inferred. This could possibly be leveraged to do automatic drone swarm detection or interference mitigation. To this end, it will be important to understand how well the proposed model matches real second-order characteristics of drone swarms and if extensions to the model are needed, such as the inclusion of noise, macro Doppler, or occlusion effects.}



\appendix[Derivation of series coefficients]\label{app:Derivation_of_series_coefficients}
To carry out the expectation $\E_\phi[J_n(l\text{sin}(\phi))]$ we expand the Bessel function using the Bessel integral and the resulting exponential using the Jacobi-Anger expansion,
\begin{equation}
    \E_\phi[J_n(l\text{sin}(\phi))] = \frac{1}{2\pi (i)^n}\int_0^{2\pi} J_0(l\text{cos}(\theta))e^{in\theta}d\theta.
\end{equation}
We recognize this as the Fourier coefficients for the even function $J_0(l\text{cos}(\theta))$, hence
\begin{equation}
    C_n(l) = \begin{cases}
\int_0^{2\pi}J_0(l\text{cos}(\theta))\text{cos}(n\theta) d\theta,& $\text{$n$ even}$,\\
0, & \text{otherwise.}\\
\end{cases}
\end{equation}
For even $n$, the integral is evaluated as,
\begin{equation}
    \int_0^{2\pi}J_0(l\text{cos}(\theta))\text{cos}(n\theta) d\theta = J_n^2\left(\frac{l}{2}\right).
\end{equation}

\bibliographystyle{IEEEtran}
\bibliography{ARM_lib}

\end{document}